\documentclass[fleqn,12pt,twoside]{article}
\usepackage{espcrc1}

\usepackage{epsfig}
\usepackage[figuresright]{rotating}

\newcommand{\AmS}{{\protect\the\textfont2
  A\kern-.1667em\lower.5ex\hbox{M}\kern-.125emS}}

\title{Dileptons and Photons from Coarse--Grained Microscopic Dynamics and 
Hydrodynamics Compared to Experimental Data}

\author{Joseph Kapusta\footnote{This work was done in collaboration with Mohamed 
Belkacem, Paul Ellis and Pasi Huovinen \cite{us}, and was supported by the US 
Department of Energy grant DE-FG02-87ER40328.}
\address{School of Physics and Astronomy, University of Minnesota\\
Minneapolis, Minnesota 55455, USA}}
       
\begin{document}

\maketitle

\begin{abstract}
We compute the radiation of dileptons and photons using relativistic
hydrodynamics and a coarse--grained version of the microscopic event
generator UrQMD, both of which provide a good description of the
hadron spectra.  The currently most accurate dilepton and photon
emission rates from perturbative QCD and from experimentally-based
hadronic calculations are used.  Comparisons are made to data on
central Pb-Pb and Pb-Au collisions taken at the CERN SPS at a beam
energy of 158 A GeV. Both hydrodynamics and UrQMD provide very good
descriptions of the photon transverse momentum spectrum measured
between 1 and 4 GeV, but very slightly underestimate the low mass spectrum
of $e^+e^-$ pairs, even with greatly broadened $\rho$ and $\omega$
vector mesons. 
\end{abstract}

\section{Introduction}

We focus on central Pb-Pb and Pb-Au collisions at 158 GeV per nucleon at the 
CERN SPS.  A coarse--graining of 100 UrQMD simulation events was performed.  
This gives flow fields of temperature, baryon chemical potential, and fluid 
velocity that can be compared to the flow fields of relativistic hydrodynamics.  
The initial conditions and final freezeout temperature in the hydrodynamic 
calculations were previously adjusted to represent the rapidity and transverse 
momentum distributions of hadrons so there are no new parameters introduced 
here.  The coarse--grained version of UrQMD does require a freezeout 
temperature; a value of 125$\pm$10 MeV reproduces not only the rapidity 
distribution very well (shown) but also the transverse momentum distributions of 
negative hadrons and the net proton number (not shown).  We then compare to the 
measured photon spectrum by summing both the hard perturbative QCD photons 
together with the thermal emission.
We also compare to the $e^+e^-$ spectrum as measured at the 
highest multiplicities.  The background cocktail is the same as used by the 
experiment itself.  The thermal rates are those computed by Eletsky, Belkacem, 
Ellis and Kapusta \cite{eletsky} based on kinetic theory and hadronic data.  
These rates are nearly identical to those computed by Rapp and Wambach 
\cite{rapp}; both of them show a significant shift in spectral weight from the 
rho-meson region to lower masses.  The results are shown in the following ten 
figures.  Further details may be found in ref. \cite{us}.    

\twocolumn

\begin{figure}[t]
  \begin{center}
    \epsfysize 68mm  \epsfbox{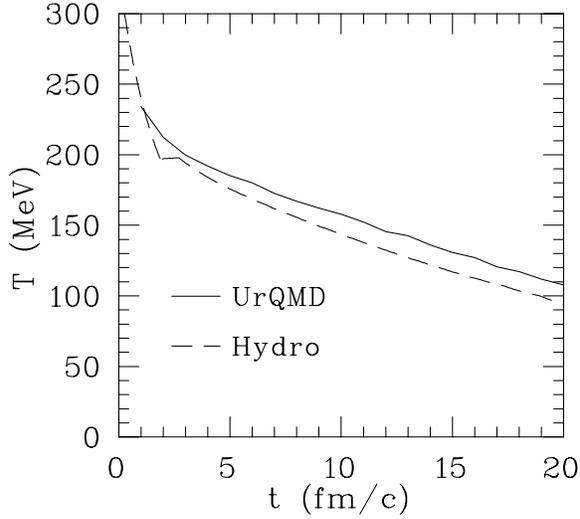}
  \end{center}
\vspace*{-12mm}
\caption{The temperature as a function of local time at 
the origin in the UrQMD model and the hydrodynamic model.}
 \label{figure1}
\end{figure}
\begin{figure}[b]
  \begin{center}
    \epsfysize 68mm \epsfbox{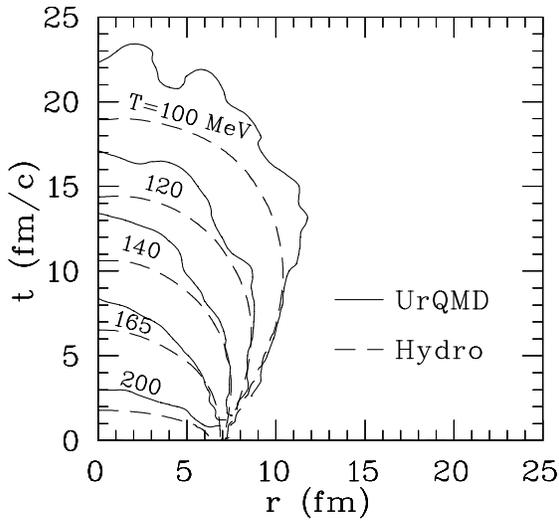}
  \end{center}
\vspace*{-12mm}
\caption{Temperature contours in local time
          and cylindrical radius in the central transverse plane for the UrQMD
          model and the hydrodynamic model.}
 \label{figure2}
\end{figure}

\begin{figure}[t]
  \begin{center}
    \epsfysize 68mm  \epsfbox{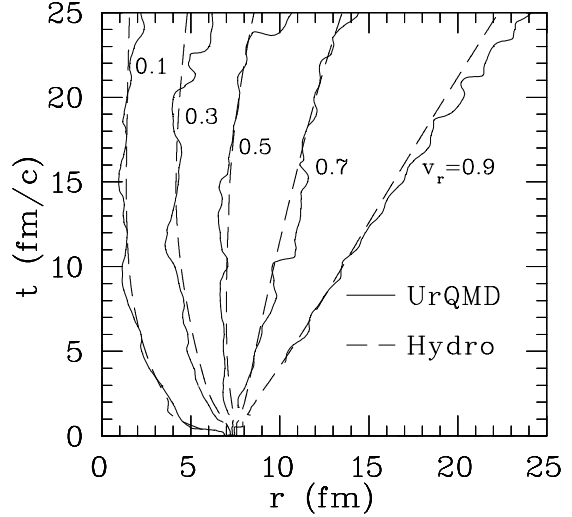}
  \end{center}
\vspace*{-12mm}
\caption{Contours of equal radial flow velocity as a function of local time
         and cylindrical radius in the central transverse plane for the UrQMD
         model and the hydrodynamic model.}
 \label{figure3}
\end{figure}
\begin{figure}[b]
  \begin{center}
    \epsfxsize 67mm  \epsfbox{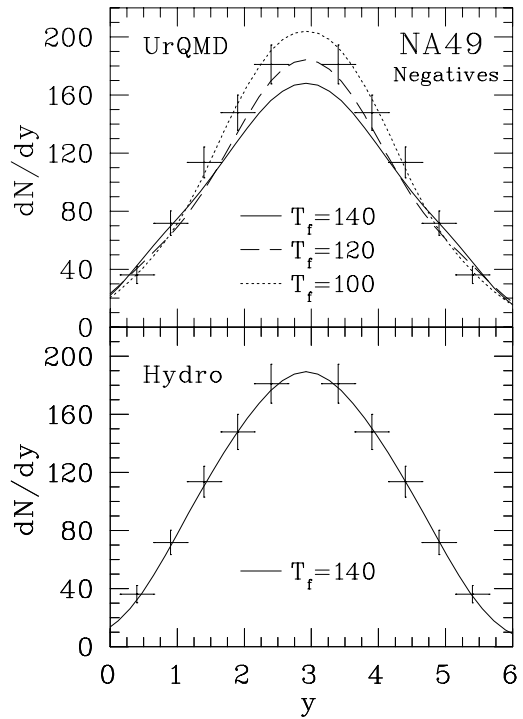}
  \end{center}
\vspace*{-12mm}
\caption{The rapidity distribution of negatively charged hadrons. Upper 
         panel: UrQMD for various freeze-out temperatures; lower
         panel: hydrodynamic model; data: NA49 collaboration
         \protect\cite{Jones}.}
 \label{figure4}
\end{figure}

\begin{figure}[t]
  \begin{center}
    \epsfysize 60mm  \epsfbox{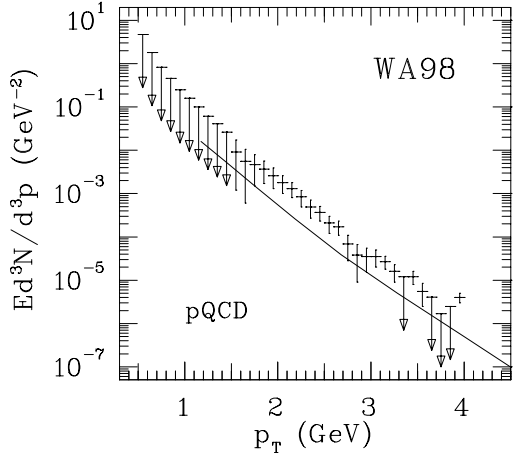}
  \end{center}
\vspace*{-12mm}
\caption{Photon spectrum from Pb-Pb collisions at 158 A GeV by the WA98 
         collaboration \protect\cite{wa98} compared to a perturbative
         QCD calculation.}
 \label{figure5}
\end{figure}
\begin{figure}[b]
  \begin{center}
    \epsfxsize 67mm  \epsfbox{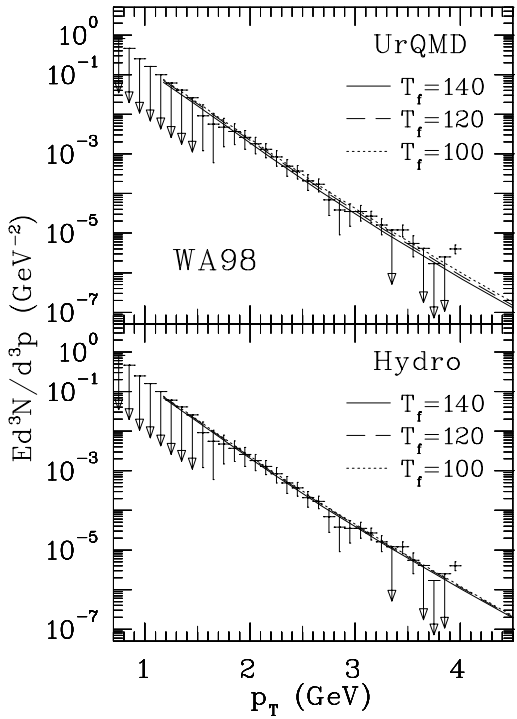}
  \end{center}
\vspace*{-12mm}
\caption{Comparison of the WA98 photon spectrum \protect\cite{wa98} to the
         predictions of the UrQMD model and the
         hydrodynamic model at several freeze-out
         temperatures.}
 \label{figure6}
\end{figure}

%
%

\begin{figure}[t]
  \begin{center}
    \epsfxsize 67mm  \epsfbox{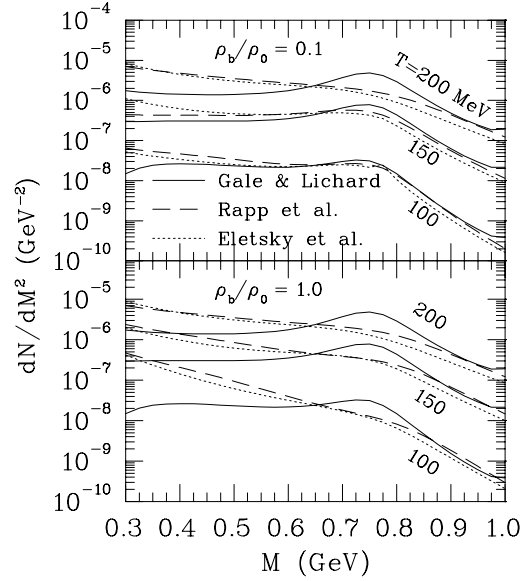}
  \end{center}
\vspace*{-12mm}
\caption{Thermal dilepton emission rates computed by Gale and Lichard
         \protect\cite{galel}, Rapp {\it et al.}
         \protect\cite{rapp} and Eletsky {\it et al.} 
         \protect\cite{eletsky} at various
         temperatures.  The baryon densities are fixed at 1/10
         and 1 times the equilibrium density of 
         cold nuclear matter.}
 \label{figure7}
\end{figure}
\begin{figure}[b]
  \begin{center}
    \epsfysize 60mm  \epsfbox{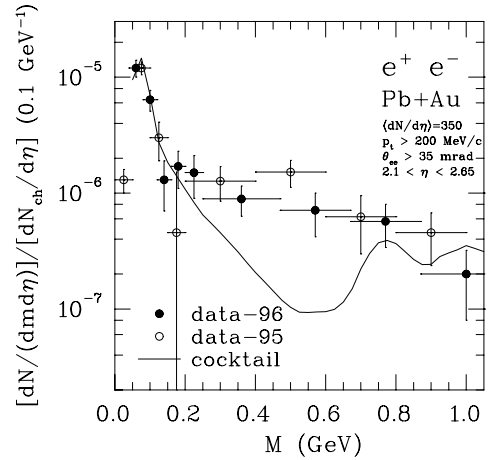}
  \end{center}
\vspace*{-12mm}
\caption{Comparison of the dilepton data for Pb-Au collisions at 158 A 
         GeV ('95 data Ref. \protect\cite{voi}, '96 data
         Ref. \protect\cite{len}) with the contribution from the decay
         of hadrons after freezeout.}
 \label{figure8}
\end{figure}

\onecolumn

%
%

\begin{figure}[t]
\begin{minipage}[t]{77mm}
  \begin{center}
    \epsfxsize 67mm  \epsfbox{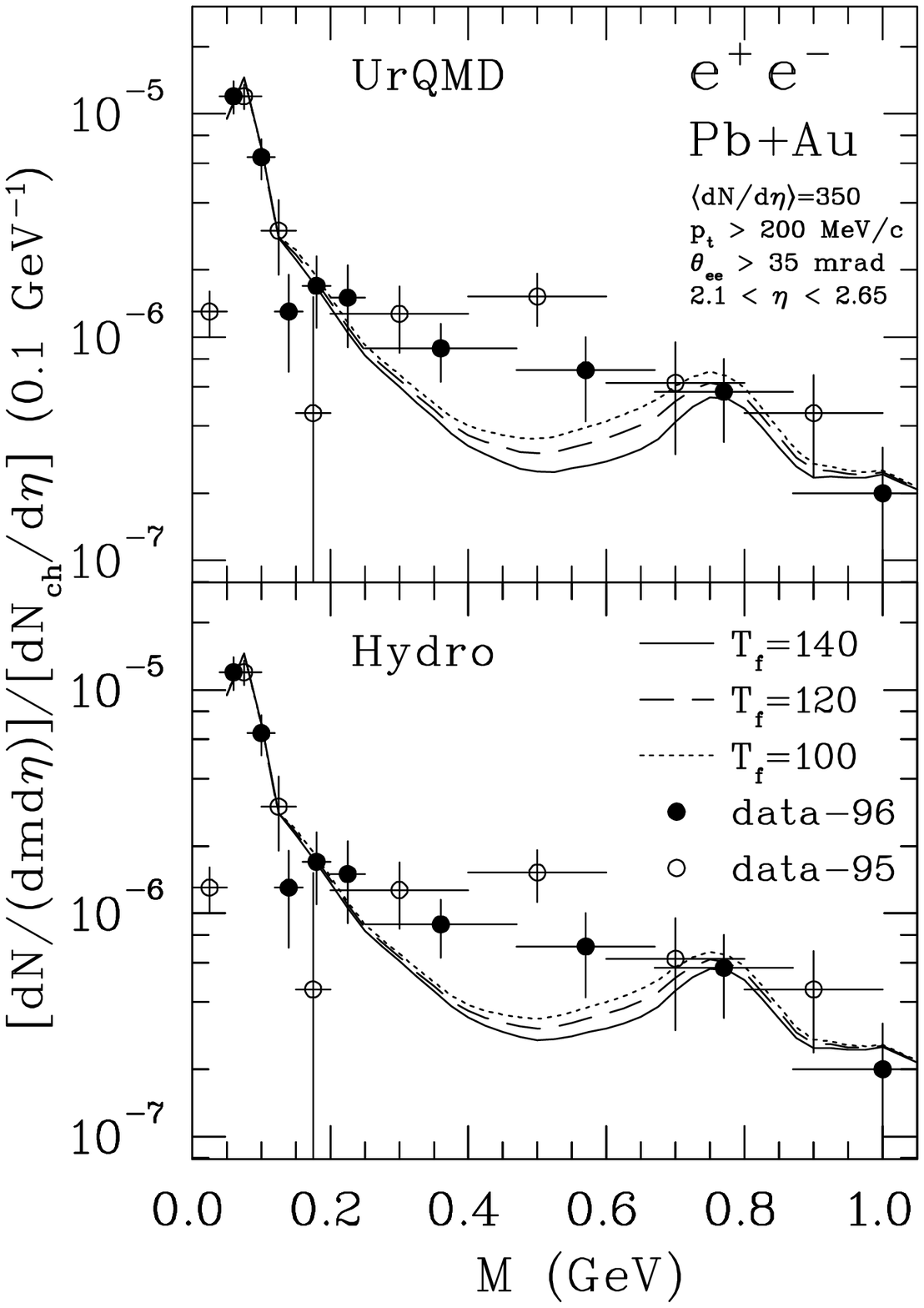}
  \end{center}
\vspace*{-12mm}
\caption{Comparison of the dilepton data \protect\cite{voi,len} with 
         predictions of the UrQMD model and the
         hydrodynamic model at several freeze-out
         temperatures.}
 \label{figure9}
\end{minipage}
\hspace{\fill}
\begin{minipage}[t]{77mm}
  \begin{center}
    \epsfxsize 67mm  \epsfbox{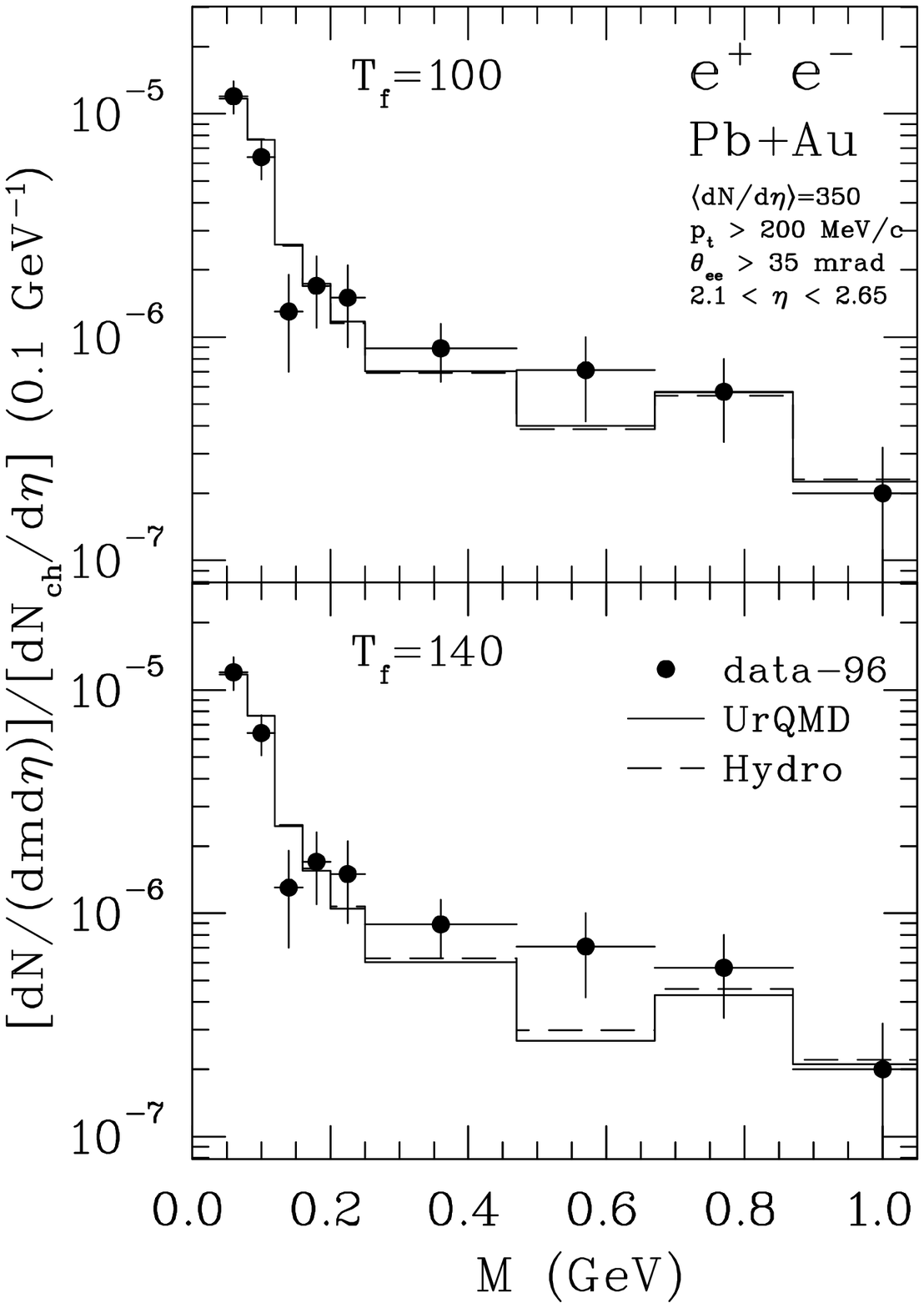}
  \end{center}
\vspace*{-12mm}
\caption{Comparison of the dilepton data \protect\cite{len} with 
         binned predictions of the UrQMD model and 
         the hydrodynamic model at two freeze-out
         temperatures.}
 \label{figure10}
\end{minipage}
\end{figure}

%
%

\section{Conclusions}

The space-time evolution of coarse--grained UrQMD is very similar to that of 
hydrodynamics.  Both represent the hadronic data quite well.  The 
thermal photon and dilepton rates are relatively well understood.  The 
theoretical predictions for both photons and dileptons agree rather well with 
data for central heavy ion collisions at 158 GeV per nucleon, although there may 
be a slight underestimate of dileptons in the mass range from 500 to 700 MeV.  
We should celebrate the success of both theory and experiment!

\end{document}